\documentclass[a4paper,11pt]{article}
\usepackage[final]{epsfig}
\usepackage{cite,subfigure}
\begin{document}
\title{\textbf{
Alignment and alignment dynamics 
of nematic liquid crystals on 
Langmuir-Blodgett mono-layers
}}
\author{
V. S. U. Fazio, L. Komitov and S. T. Lagerwall\\
\textit{\small{Department of 
Microelectronics \& Nanoscience}}, \\
\textit{\small{Chalmers University of Technology
\& G\"oteborg University,}}\\
\textit{\small{S-41296 G\"oteborg, Sweden}}
}
\date{}
\maketitle
%
%\clearpage
%
\begin{abstract}
\noindent
Mono-layers of stearic and  behenic acids deposited 
with the Langmuir-Blodgett technique,
were used as aligning films in nematic liquid crystal 
cells.
During the filling
process the liquid crystal
adopts a deformed quasi-planar alignment with splay-bend deformation
and preferred orientation along the filling direction.
This state is metastable and transforms with time into homeotropic
once the flow has ceased.
The transition is
accompanied by formation of disclination lines which nucleate
at the edges of the cell.
The lifetime of the metastable splay-bend state was found to depend on the
cell thickness.  
On heating, anchoring transition from quasi-homeotropic to degenerate
tilted alignment in form of circular domains takes place near the
transition to the isotropic phase.
The anchoring transition is reversible with a small hysteresis.
\end{abstract}
%
%
%
%\clearpage
%
%\psdraft
%
%
%
\section{Introduction}
Liquid crystal cells exhibiting uniform orientational alignment 
over large areas are required in most device applications.
The usual techniques used to obtain a preferred orientation,
or anchoring direction, relay on inducing physical or chemical interactions
between a prepared substrate and the liquid
crystal molecules\cite{Jerome91}.
Among them the Langmuir-Blodgett (\textsf{LB}) 
technique,
%\cite{Langmuir17,Blodgett35},
which enables the deposition of organic aligning films with controlled
molecular order and thickness and very high 
reproducibility\cite{Petty,Gaines},
is still in a research state but represents a high potential for
future display alignment in an industrial scale.

When a substrate has been covered with a \textsf{LB} mono-layer,
a possible aligning mechanism
for a liquid crystal is that the molecules penetrate into the
layer of aliphatic chains.
They then adopt the orientation of these chains, which leads to
homeotropic or conical anchoring, depending on the chains' 
orientation\cite{Jerome91}.

In this work mono-layers of stearic (\textsf{C18}) and behenic
(\textsf{C22}) acids were used as aligning layers in 
nematic liquid crystal (\textsf{NLC}) cells for
obtaining a uniform homeotropic 
surface-induced orientation.\cite{KomSteGabPug94,SauStaSmiDan85}
The alignment was studied during the filling process and pursued during the
relaxation to the equilibrium state.
The temperature stability of this state was also investigated.

\section{Experiment}
\subsection{Film preparation}
Mono-layer formation was achieved by spreading a solution of
stearic or behenic acid (see Table \ref{fattyacids})
on the surface of ultrapure
Milli-Q water in a \textsf{LB} trough (\textsf{KSV 3000}) held in a clean 
room environment to limit contamination of the trough by dust.
\begin{table}
\begin{center}
\caption{
\small{
Long-chain fatty acid compounds used in this experiment.
}}
\begin{tabular}{ccc}
\vspace{0.5mm}\\
\hline
\hline
\hspace{2mm} Common name \hspace{2mm} & 
\hspace{2mm} Structure \hspace{2mm} & 
\hspace{2mm} Abbreviation \hspace{2mm}  \\
\hline
stearic & C$_{17}$H$_{35}$COOH & \textsf{C18}\\
behenic & C$_{21}$H$_{43}$COOH & \textsf{C22}\\
\hline
\hline
\end{tabular}
\label{fattyacids}
\end{center}
\end{table}

The substrate used in the experiments was 
tin oxide (\textsf{ITO}) coated glass.
%\textit{with a resistivity in the range of ...., supplied by ....}.
Glass plates were cut to a size of 75\,mm$\times$30\,mm
and pre-cleaned in an ultrasound bath filled with ultrapure water
to eliminate the bigger dust particles and the
residues from cutting.
They were then cleaned during 8 minutes
in an ultrasound bath filled with a mixture of
5 parts of H$_{2}$O, 1 part of NH$_{3}$, and 1 part of 
H$_{2}$O$_{2}$ at a temperature of 80$^{\circ}$C.
Finally they were rinsed in a three stage cascade with Milli-Q water
and again rinsed and dried in a centrifuge.
With this procedure we could eliminate both organic and
inorganic contaminations.
The glass plates were stored in the clean room so that they 
could last cleaned for at least two weeks.

The stearic and behenic solutions were made to
a concentration of 1\,mM in Merck chloroform.
The sample material was spread on the water surface with a clean,
all glass, Hamilton syringe.
After a time lapse of ca.\,10\,minutes to allow the solvent to
evaporate, the mono-layer was compressed at a a rate of 
approximately 0.3$\times$10$^{-3}$\,(nm$^{2}$\,s$^{-1}$molecule$^{-1}$) 
until the
requested pressure was reached.

The glass substrates were immersed in the sub-phase before
spreading the mono-layer, and the transfer to the glass occurred during 
the extraction of the glass from the sub-phase at a controlled
speed (10\,mm/min), keeping the surface pressure constant.

\subsection{Cell fabrication and observations}
\begin{figure}
\begin{minipage}{\textwidth}
\parbox[b]{0.5\textwidth}{
\epsfig{file=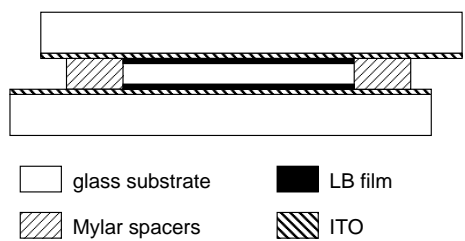, width=0.48\textwidth}}
\hfill
\parbox[b]{0.49\textwidth}{
\protect\caption{\small{
Cross-section of a cell. The \textsf{ITO} layer is in the
inner part of the cell with the \textsf{LB} film on top of it.
The proportions are symbolic.
}}
{\label{cells}\sloppy }
}
\end{minipage}
\end{figure}

Sandwich cells were made with the \textsf{ITO} layers in the
inner part (Figure \ref{cells}) and spaced 
with polyester
of various thickness supplied by Mylar.
Care was taken not  to touch, and thus contaminate, the inside
surface of the cells.

The cells were capillary filled with \textsf{MBBA} 
(supplied by Aldrich) at room temperature.
The observations on them were made with a microscope where 
the sample is inserted in a hot stage
between crossed polarisers.
Due to the birefringence of liquid crystals the planar phase,
where the molecules lie parallel to the planes of the polarisers, appears very
coloured and changes appearance as the sample is rotated.
On the other hand, the homeotropic phase, where the long axis of
the liquid crystal is oriented perpendicularly to the planes of the
polarisers, looks uniformally dark.
The microscope was also equipped with a video-camera connected to
a computer: it was possible to  
take pictures of the 
samples at regular time intervals and then 
record the evolution of the alignment state.

\section{Results}

\subsection{Isotherms}
\begin{figure}
\begin{center}
\epsfig{file=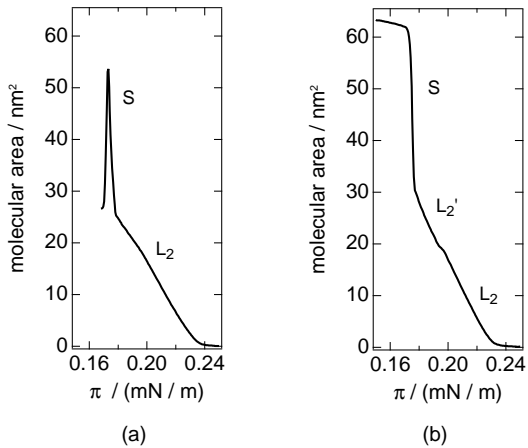,width=0.6\textwidth}
\caption{\small{
\label{isoterme_single}
(a) Surface pressure versus area per molecule isotherm for
stearic acid. 
We can distinguish the \textsf{L}$_{2}$,
or liquid-condensed  phase\protect{\cite{Petty}}, and the \textsf{S} or solid
phase.  
(b) Surface pressure versus area per molecule isotherm for
behenic acid.
In addition to the \textsf{L}$_{2}$ and the \textsf{S} phases,
here we have a slightly different liquid-condensed phase, the
\textsf{L'}$_{2}$ phase.
See Table \protect{\ref{monolayerphases}} for the
phase characteristics.
}}
\end{center}
\end{figure}

In Figure \ref{isoterme_single} the surface pressure versus
molecular area isotherms for films of stearic 
acid (\textsf{C18}) and behenic acid (\textsf{C22})
are shown.
Table \ref{monolayerphases} provides a summary of the three
mono-layer phases observed in the isotherms of Figure \ref{isoterme_single}.
\begin{table}[b]
\caption{\small{
\label{monolayerphases}
Condensed mono-layer phases for fatty acids. (After Petty, 1996.)
}}
\begin{tabular}{ccl}
\vspace{0.5mm}\\
\hline
\hline
Phase & Name & Characteristics \\
\hline
\textsf{L}$_{2}$ & liquid-condensed &
 Slightly tilted molecular chains. \\
\textsf{L'}$_{2}$ & liquid-condensed & 
 Tilted chains, but with tilt direction in excess\\
& & of 45$^{\circ}$ relative to 
 \textsf{L}$_{2}$ phase; similar compressibi-\\
& & lity as \textsf{L}$_{2}$ phase. \\
\textsf{S} & solid & Upright molecules; 
less compressible than \textsf{L}$_{2}$\\
& & and \textsf{L'}$_{2}$ phases;
high collapse pressure.\\
\hline
\hline
\end{tabular}
\end{table}

The deposition pressure for the \textsf{LB} films was chosen at
20 mN/m.
In this condition we have the 
liquid-condensed phase of the mono-layer for both
compounds (see Figure \ref{isoterme_single}).

\subsection{Alignment}
\begin{figure}
\subfigure[
\label{cellfilling}
]
{\epsfig{file=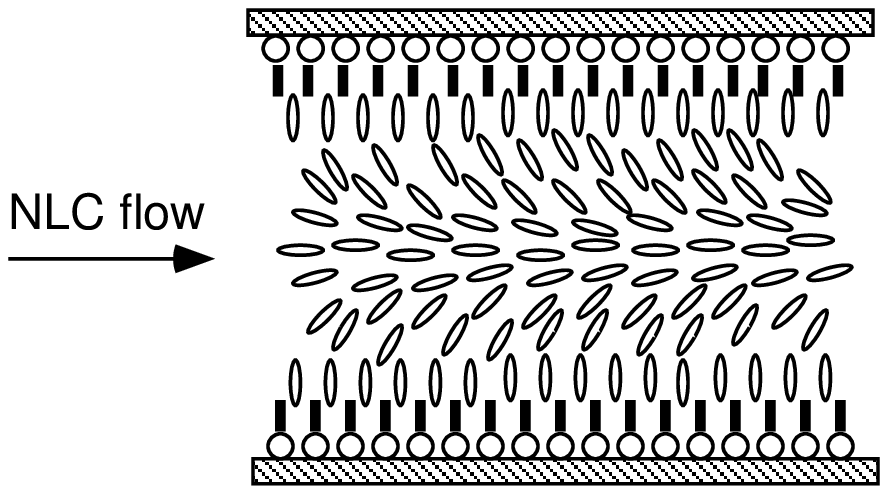,width=0.48\textwidth}}
\subfigure[
\label{conoscoplanar}
]
{\epsfig{file=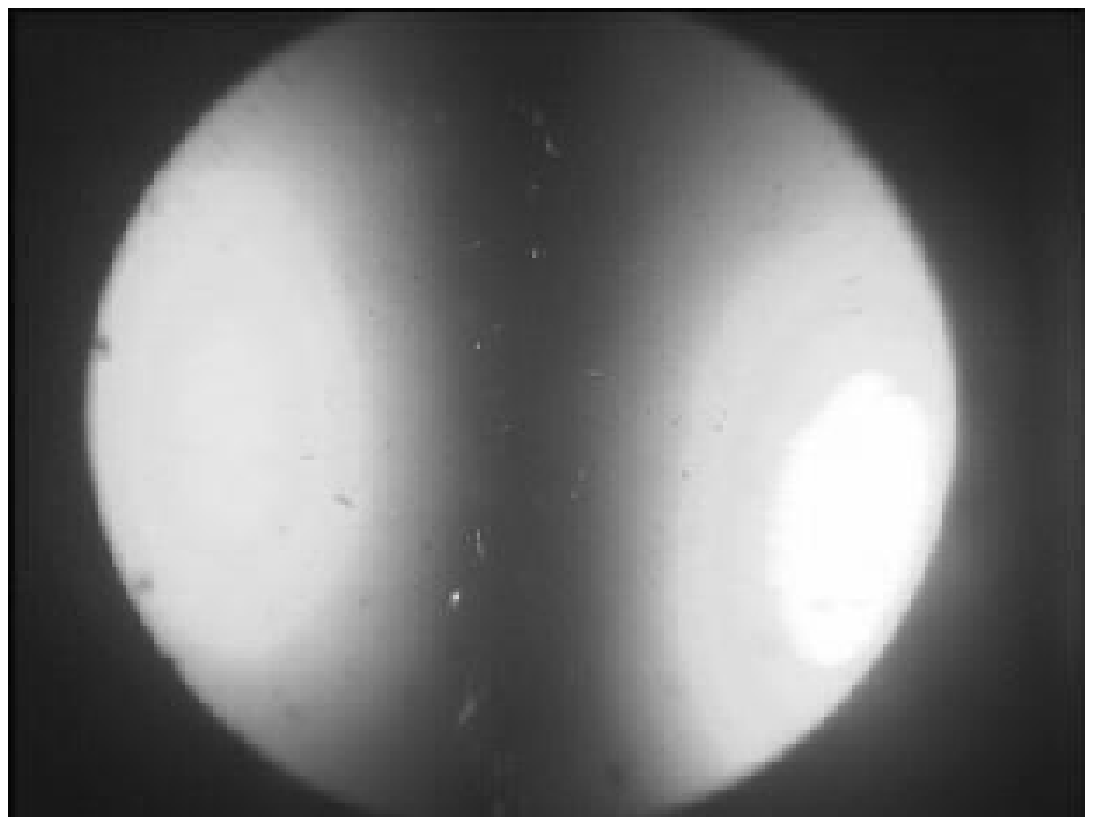,width=0.49\textwidth}}
\caption{
\small{(a) During filling the orientation of the \textsf{LC}
molecules is quasi-planar 
(the molecules lie parallel to the glass substrate)
with a preferred  alignment along
the filling direction.
The molecules in the centre of the cell are essentially parallel to the
substrate and a splay-bend deformation in the \textsf{NLC} is induced
by the presence of the aligning layer{\protect\cite{KomLagSpaSte92}}.
(b) Conoscopic picture of a 22.3\,$\mu$m thick cell with aligning
\textsf{C18} mono-layer during the \textsf{NLC} filling.
The picture shows that the \textsf{NLC} is oriented
as depicted in Figure 
(a){\protect\cite{Ehlers}}: 
the molecules in the center of the cell are aligned in the
filling direction and a splay-bend structure is induced 
by the presence of the aligning \textsf{LB} films{\protect\cite{KomHauKos86}}.
}}
\end{figure}

The cells were capillary filled with \textsf{MBBA} at room temperature
(where \textsf{MBBA} is in the nematic phase).
During filling 
we have the alignment condition in which
the orientation of the \textsf{MBBA} molecules is 
quasi-planar with a preferred alignment 
along the filling direction:
the molecules in the centre of the cell are essentially parallel to the
substrate and a splay-bend deformation in the \textsf{NLC} is induced
by the presence of the aligning layer\cite{KomLagSpaSte92} 
(see Figures \ref{cellfilling}
and \ref{conoscoplanar}).
%
%\psdraft
%\psfull
%
As soon as the flow stops, because the cell is completely filled with 
\textsf{MBBA}, domains of homeotropic alignment start to nucleate at the
edges of the sample and continuously grow until the whole
sample becomes homeotropic.

\begin{figure}
\begin{center}
\epsfig{file=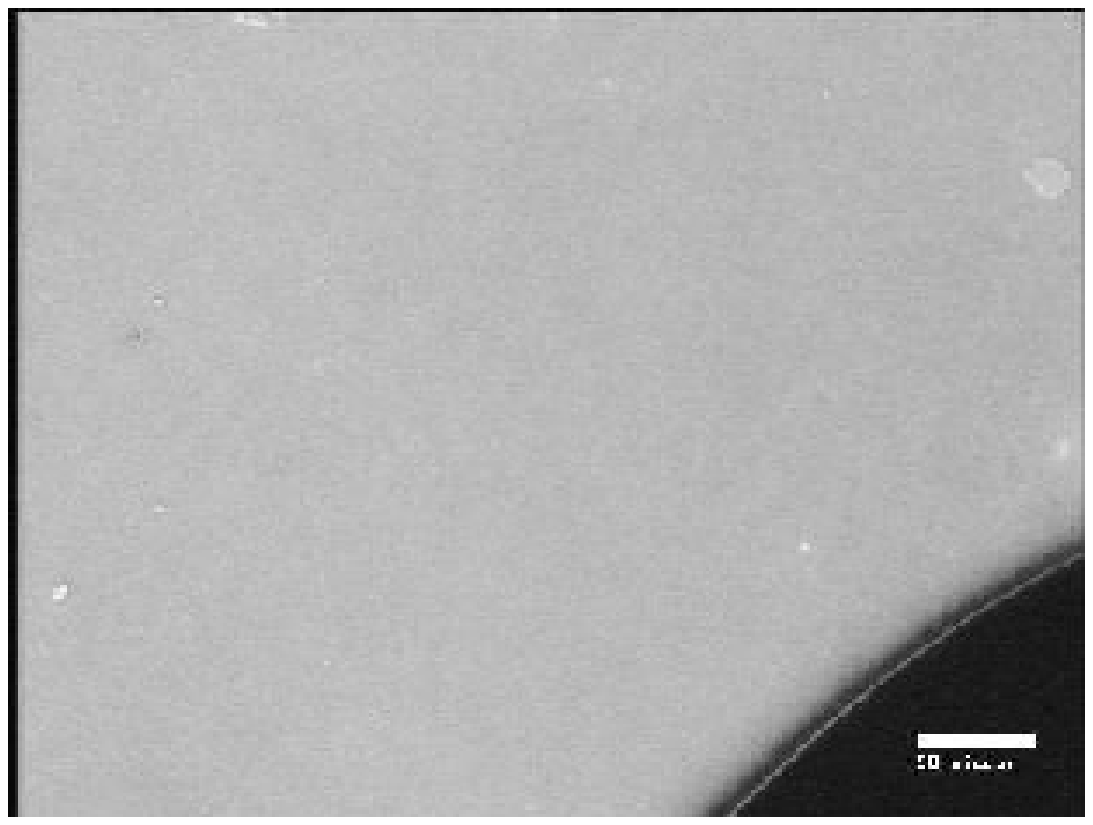,width=0.48\textwidth}
\epsfig{file=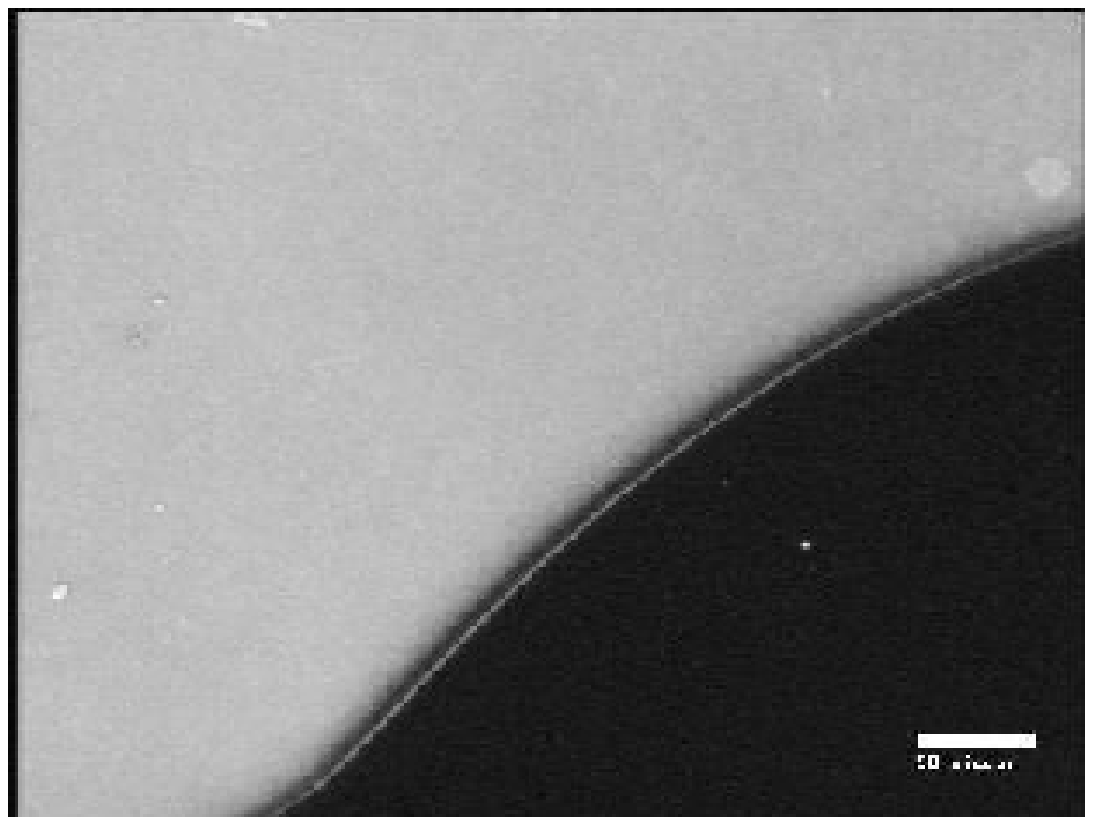,width=0.48\textwidth}
\caption{\small{
\label{expand}
Cell between crossed polarisers.
The \textsf{LB} aligning mono-layer is \textsf{C18} and the
thickness of the cell is 12.5\,$\mu$m.
The cell is completely filled with \textsf{MBBA}, 
the flow has ceased and the homeotropic domain
(dark) expands into the quasi-planar domain (light).
The two pictures were taken with a time interval of 30\,s.
}}
\end{center}
\end{figure}

\begin{figure}
\subfigure[]
{\epsfig{file=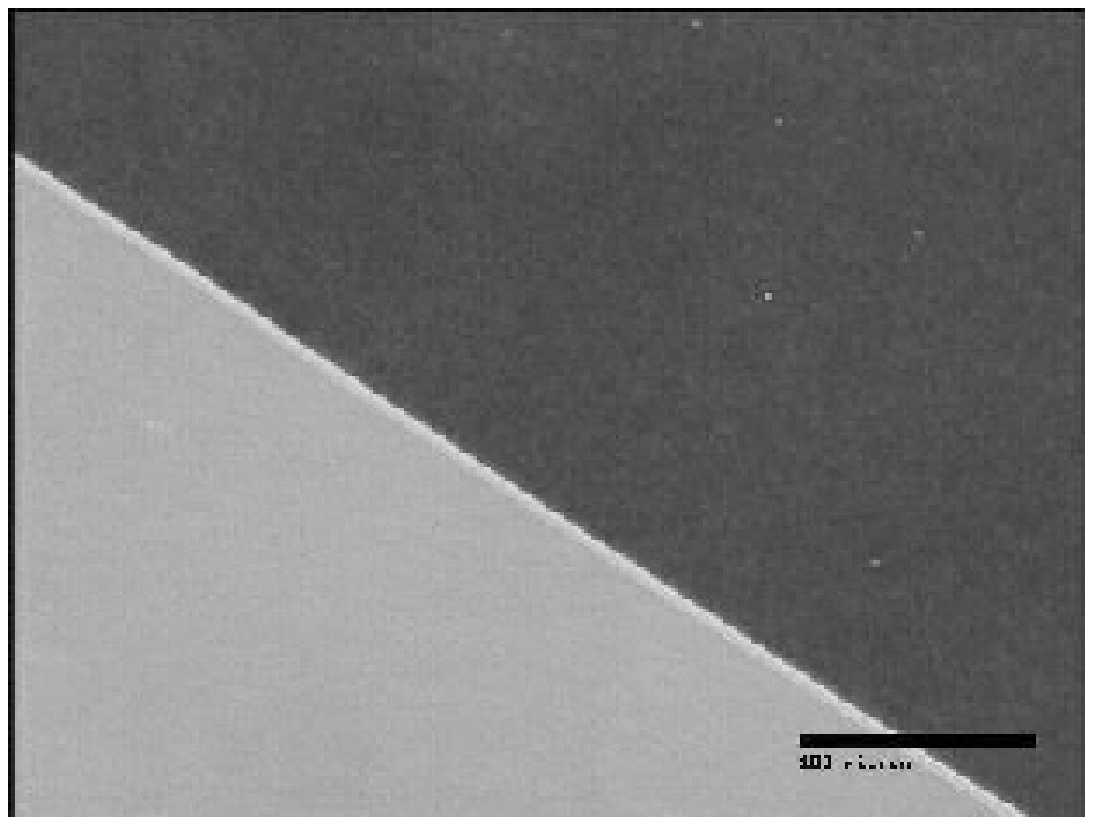,width=0.48\textwidth}}
\subfigure[]
{\epsfig{file=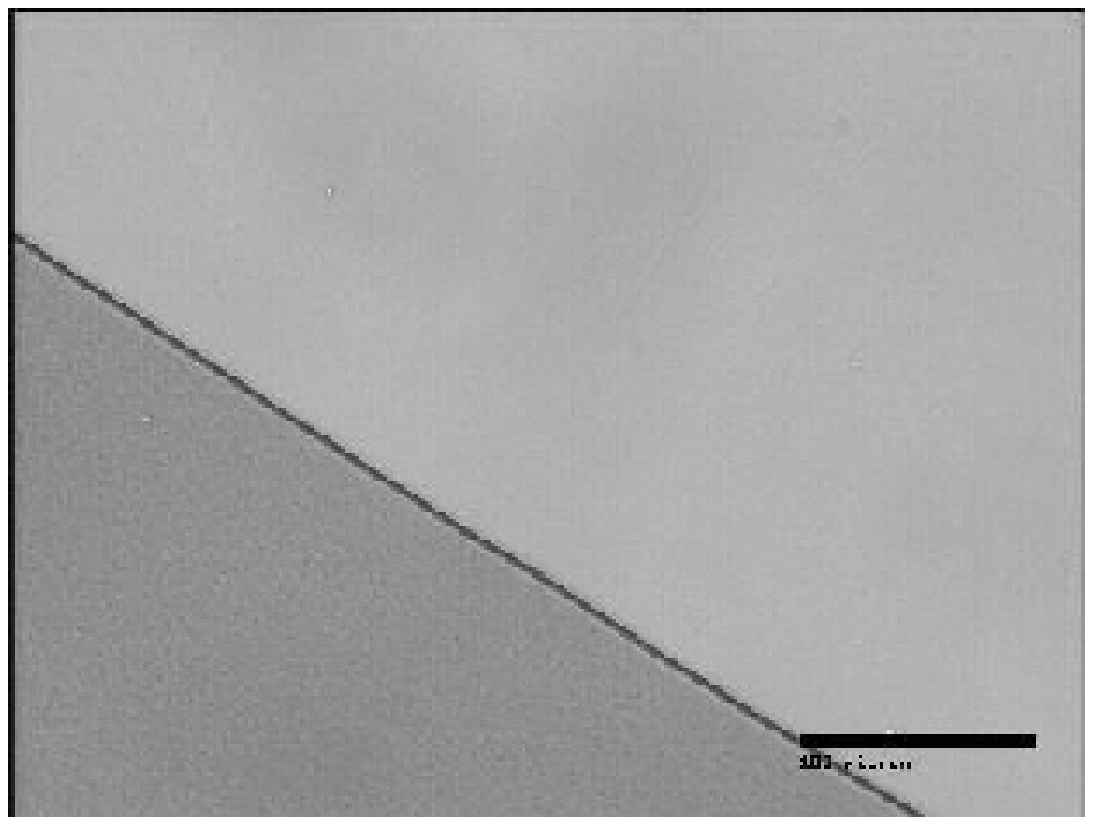,width=0.48\textwidth}}
\caption{
\label{line}
\small{
(a) Cell between crossed polarisers. 
The aligning mono-layer is \textsf{C18} and the cell thickness
is 14.4\,$\mu$m. 
The disclination line appears bright.
(b) The same cell, a few seconds later, now between parallel polarisers.
The disclination line appears dark.
}}
\end{figure}

An example of how the homeotropic domains expand in the
cell is given in Figure \ref{expand}.
%
%\psdraft
The line which devides the homeotropic domain from the quasi-planar one is
found to be a disclination line of strength $|S|=1/2$.
Such a disclination line should appear bright between crossed polarisers and
dark between parallel polarisers (see Figure \ref{line}).
%
%\psdraft
%\psfull
%

\begin{figure}
\begin{center}
\epsfig{file=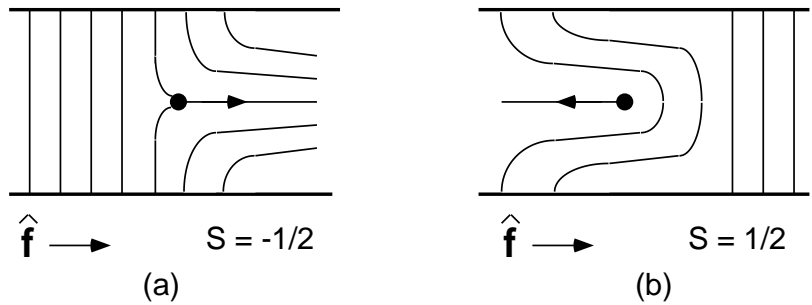,width=0.85\textwidth}
\caption{
\small{
Schematic model of disclinations of strength $|S|=1/2$.
\textbf{\^{f}} is the filling direction.
The parallel vertical lines represent the homeotropically aligned area.
The black circles represent the singularities which move into the quasi-planar 
splay-bend deformed domain.
(a) An $S=-1/2$ singularity propagates along the filling direction.
(b) An $S=1/2$ singularity propagates against the filling direction.
When two such lines meet, the singularities annihilate (-1/2 + 1/2 =0),
leaving a defect-free homeotropic domain.
}
\label{defect}
}
\end{center}
\end{figure}

A simple scheme of disclination lines
of strength $|S|=1/2$ is depicted in Figure
\ref{defect}. 
%\psfull
%
Because of the \textsf{LB} aligning layer the defect moves into the
quasi-planar area, the effect being the expansion of the homeotropic domain.

A priori, we cannot know how much the fatty acid chains are affected by the
filling flow.
At a solid surface the flow velocity is zero but here it may not be zero 
in the boundary layer of the chains.
Nevertheless we have in Figure \ref{cellfilling} depicted these chains as
being unaffected by the flow and shown them in their homeotropic equilibrium
(static) condition.
Under this hypothesis, the only driving mechanism behind the propagation of
the splay-bend into 
homeotropic alignment is the elastic distorsion in the liquid
crystal. 

\begin{figure}
\begin{center}
\epsfig{file=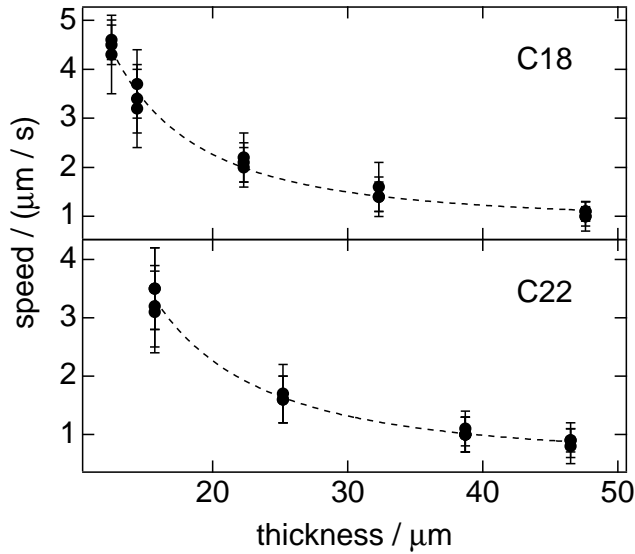,width=0.74\textwidth}
\caption{\small{
\label{speedC18C22}
Speed of expansion of the homeotropic domains as a function
of the cell thickness, for cells with \textsf{C18} and \textsf{C22}
as aligning layers.
The dots are the experimental values and the dashed lines are
fits to the function
$a + b L^{-2}$, where $a$ and $b$ are two constants.
}}
\end{center}
\end{figure}

We measured the speed with which the homeotropic domains expand
in the cells by taking pictures at fixed time intervals for
several cell thicknesses.
The speed was calculated as the area covered by the front of the
homeotropic domains in the time interval, divided by the length of
the front and the time interval.
The results are shown in Figure \ref{speedC18C22}.

For both mono-layer materials Figure \ref{speedC18C22} shows that the 
speed of expansion of the homeotropic domains decreases as the 
cell thickness increases.
For cells thinner than 12\,$\mu$m the speed of expansion of the homeotropic
domains is even larger that the actual speed of filling, so that no
relaxation process can be observed.

Whereas the propagation speed of the disclination lines depends on the layers
thickness, it does not depend on time.
It is thus not a diffusive process.
This conforms well with our previous hypothesis that the driving mechanism is
the elastic relaxation of the splay-bend  
deformation in the liquid crystal, because
the elastic torque will everywhere be the same behind the propagating front.
Its speed will therefore be directly related to the speed of relaxation.

\begin{figure}
\begin{center}
\epsfig{file=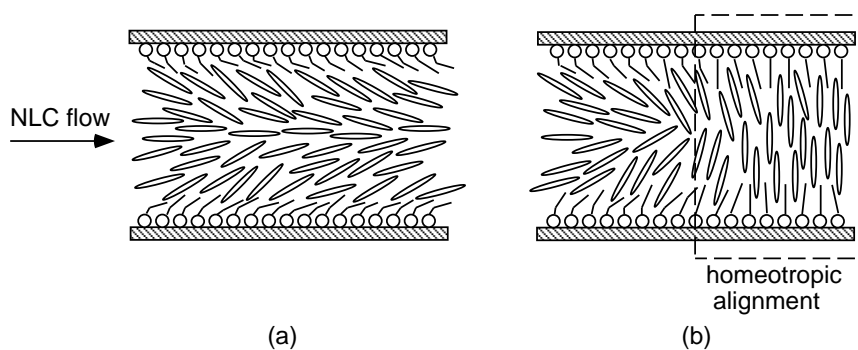,width=0.9\textwidth}
\caption{
\label{cellfilling_old}
\small{
(a) During filling the \textsf{LB} boundary layer seems to be strongly
influenced with the chains aligning along the filling direction.
(b) When the flow stops the \textsf{LB} film and the splay-bend deformed 
liquid crystal both contribute to the relaxation towards the homeotropic state.
}}
\end{center}
\end{figure}

For a small disturbance of amplitude $\delta n$ and wave vector $q$ we
may write the elastic free energy density as
\begin{equation}
\mathcal{F}=\frac{1}{2} K q^{2} (\delta n)^{2},
\label{uno}
\end{equation}
where $K$ is the characteristic elastic constant.
The elastic torque
\begin{equation}
\Gamma = -\frac{\partial \mathcal{F}}{\partial \delta n}
=-K q^{2} \delta n
\label{due}
\end{equation}
then gives the dynamic equation
\begin{equation}
\gamma \frac{d \delta n}{d t} + K q^{2} \delta n =0,
\label{tre}
\end{equation}
where $\gamma$ is the viscosity of the liquid crystal.
The characteristic time of relaxation is then
\begin{equation}
\tau = \frac{\gamma}{K q^{2}} \sim L^{2},
\label{quattro}
\end{equation}
where $L$ is the cell thickness.
In the present case this relaxation time towards 
the homeotropic state is proportional to the inverse speed
with which the homeotropic domains expand in the quasi-planar domains,
thus
\begin{equation}
v \sim L^{-2}.
\label{cinque}
\end{equation}
We fitted the experimentally measured speed of expansion of the homeotropic
domains to functional relations and found, instead (cfr. Figure
\ref{speedC18C22}), a good agreement with
\begin{equation}
v = a + b L^{-2}.
\label{sei}
\end{equation}
As $L \rightarrow \infty$, $v \rightarrow a$, and we may write
\begin{equation}
v = v_{s} + b L^{-2},
\label{sette}
\end{equation}
where $v_{s}$ is a velocity of the order of 1\,$\mu$m/s.
As $v_{s}$ is independent on $L$ we interpret it as 
characteristic of the surface,
coming from a rapid
relaxation in the boundary layer of the chains, 
which propagates the homeotropic
state into the liquid crystal bulk.
We thus have to conclude that the chains
are distorted by the filling flow, but rapidly
and forcefully relax to their equilibrium (static) state.
In other words, we believe that the \textsf{LB} film itself does have an active
r\^ole in the dynamics of the alignment transition. 
Therefore Figure \ref{cellfilling_old} probably shows a more correct
picture of the surface anchoring than Figure \ref{cellfilling} which
corresponds to our original hypothesis.

\begin{figure}
\subfigure[]{\epsfig{file=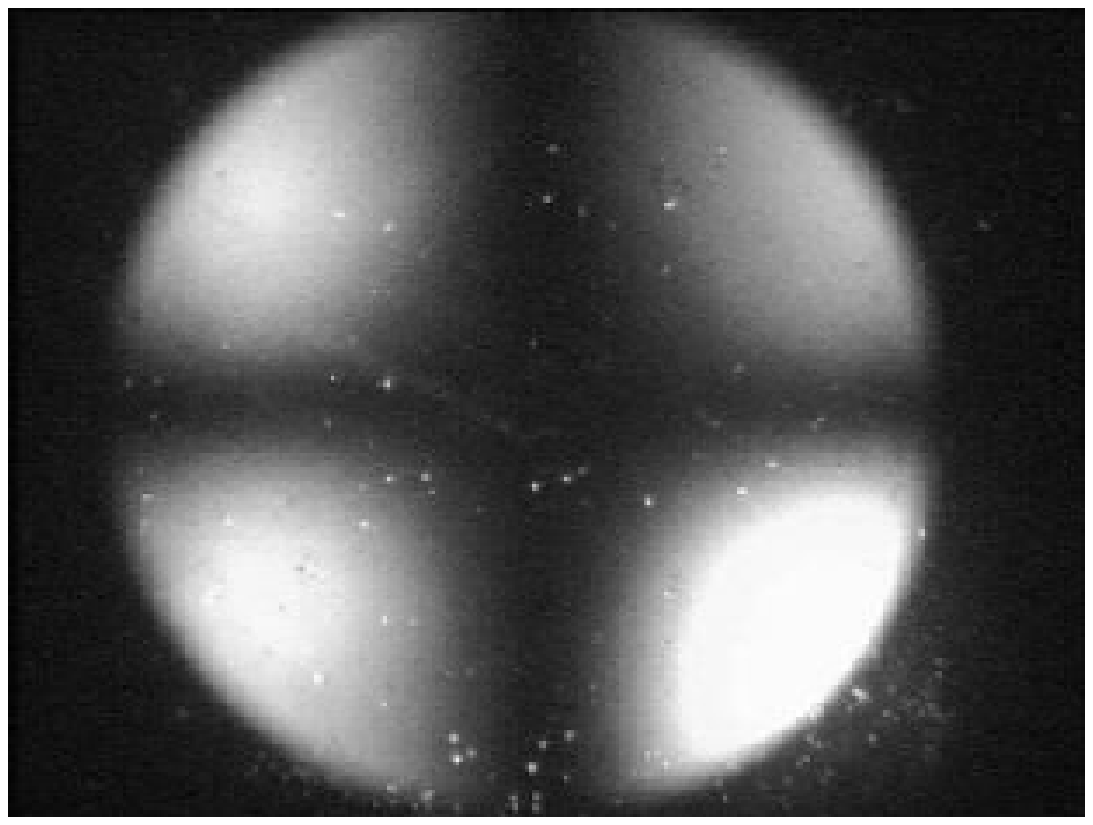,width=0.48\textwidth}}
\subfigure[]{\epsfig{file=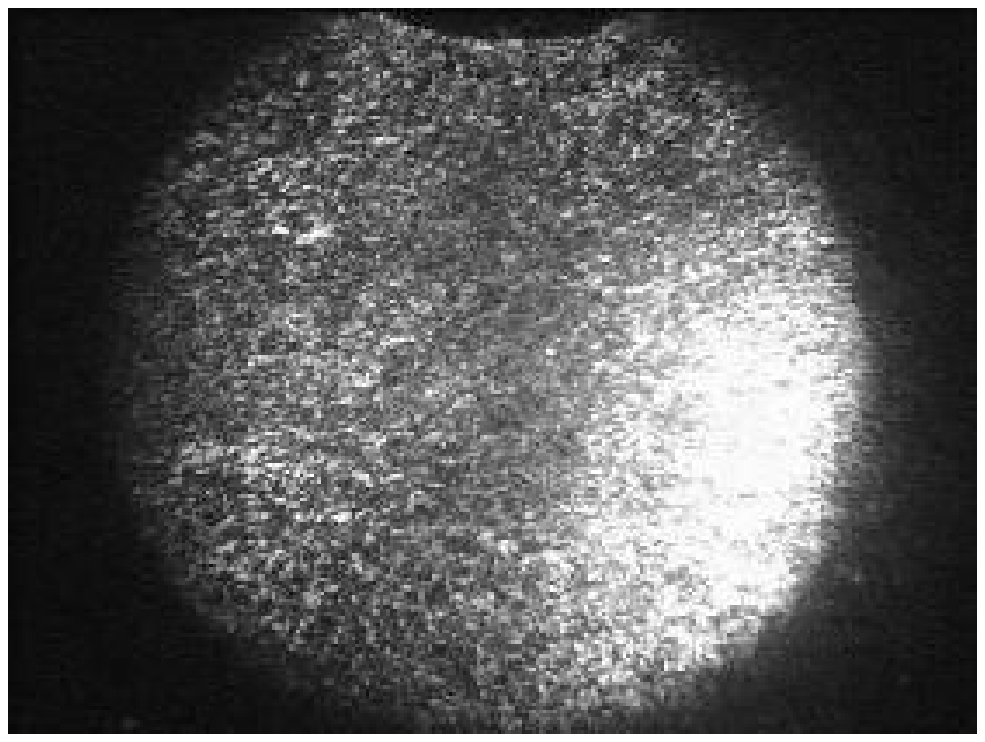,width=0.48\textwidth}}
\protect
\caption{\small{
\label{conoscopy}\sloppy
(a) Conoscopic picture of a 14.4\,$\mu$m thick cell
with \textsf{C18} as aligning layer:
a very good homeotropic alignment is achieved.
(b) Conoscopic picture of a 15.7\,$\mu$m thick cell
with \textsf{C22} as aligning layer:
the alignment is not homeotropic and not well defined.
For both mono-layer materials the alignment is found 
to be independent on cell thickness.
}}
\end{figure}

The homeotropic alignment was studied by conoscopy 
and the conoscopic pictures of samples with the two aligning layers
are shown in Figure \ref{conoscopy}.
%
%\psdraft
%\psfull
%
As we can see from the figure, a good homeotropic alignment was obtained
with \textsf{C18} as aligning layer and a much less good one with 
\textsf{C22} as aligning layer.
We believe that the reason for the different anchoring properties
should be traced to 
the behaviour of \textsf{C18} and \textsf{C22} at the air-water
interface, i.e. in the isotherms of Figure \ref{isoterme_single}.
At the deposition pressure of 20\,mN/m \textsf{C18} is in the 
liquid-condensed \textsf{L}$_{2}$ phase, while \textsf{C22} is in the
slightly more condensed liquid-condensed \textsf{L'}$_{2}$ phase,
where the molecules are strongly tilted  with respect to the
molecules in the \textsf{L}$_{2}$ phase. 
It is likely that this large tilt of the behenic acid chains 
causes a large pretilt of the \textsf{NLC} molecules 
instead of homeotropic alignment.

The alignment was found to be independent on the cell thickness,
indicating the main r\^ole of the mono-layer material.

\subsection{Anchoring transition}
On heating, we observed a first-order
anchoring transition\cite{KomSteGabPug94}
in a very narrow temperature
range, just below the clearing point.
At the transition a set of bright circular domains 
with dark crosses appear in the sample;
at constant temperature, they grow and colalesce, 
forming larger domains.
The appereance of these domains between crossed polarisers
is consistent with a degenerated tilted
orientation of the \textsf{NLC} molecules, or conical 
anchoring, also expected in the case of \textsf{LB} aligning 
films\cite{Jerome91}.

\begin{figure}
\begin{center}
\subfigure[]{\epsfig{file=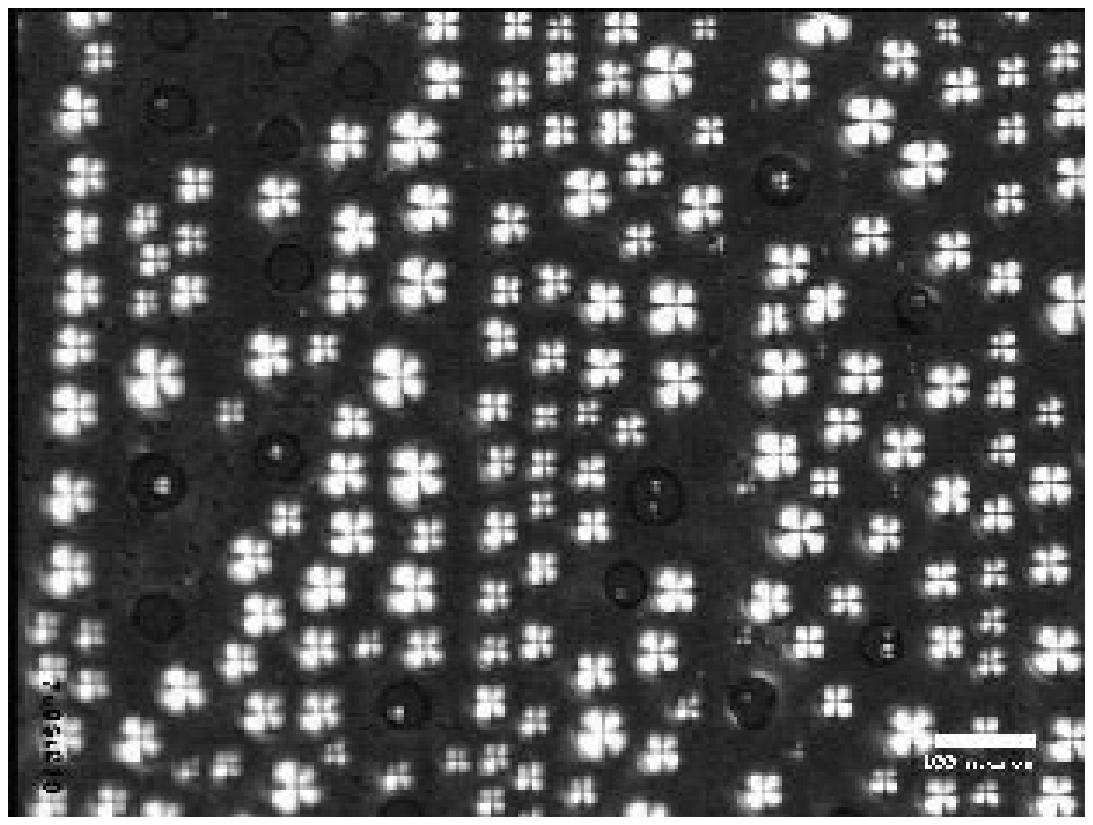,width=0.70\textwidth}}
\subfigure[]{\epsfig{file=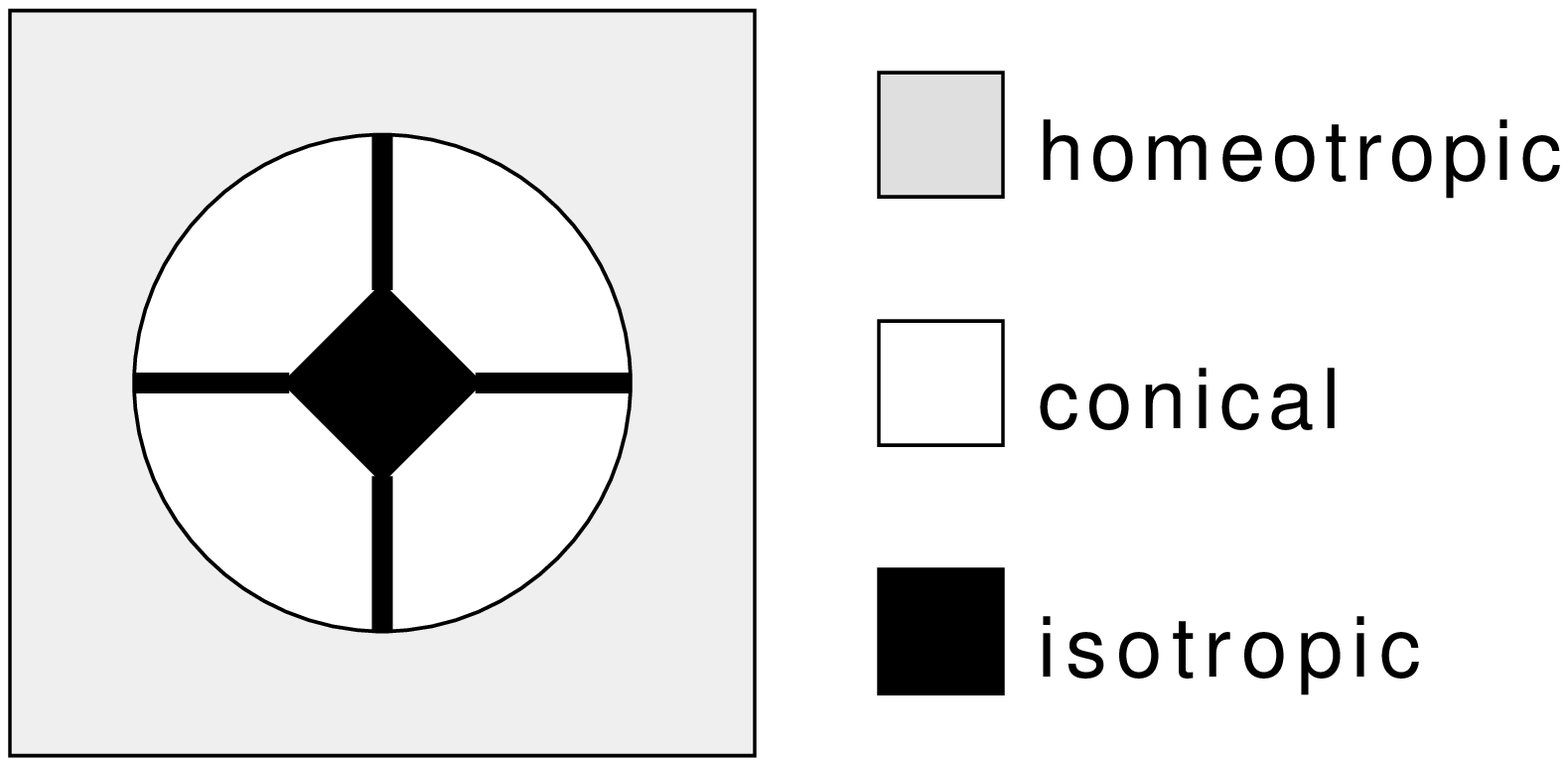,width=0.25\textwidth}}
\\
\subfigure[]{\epsfig{file=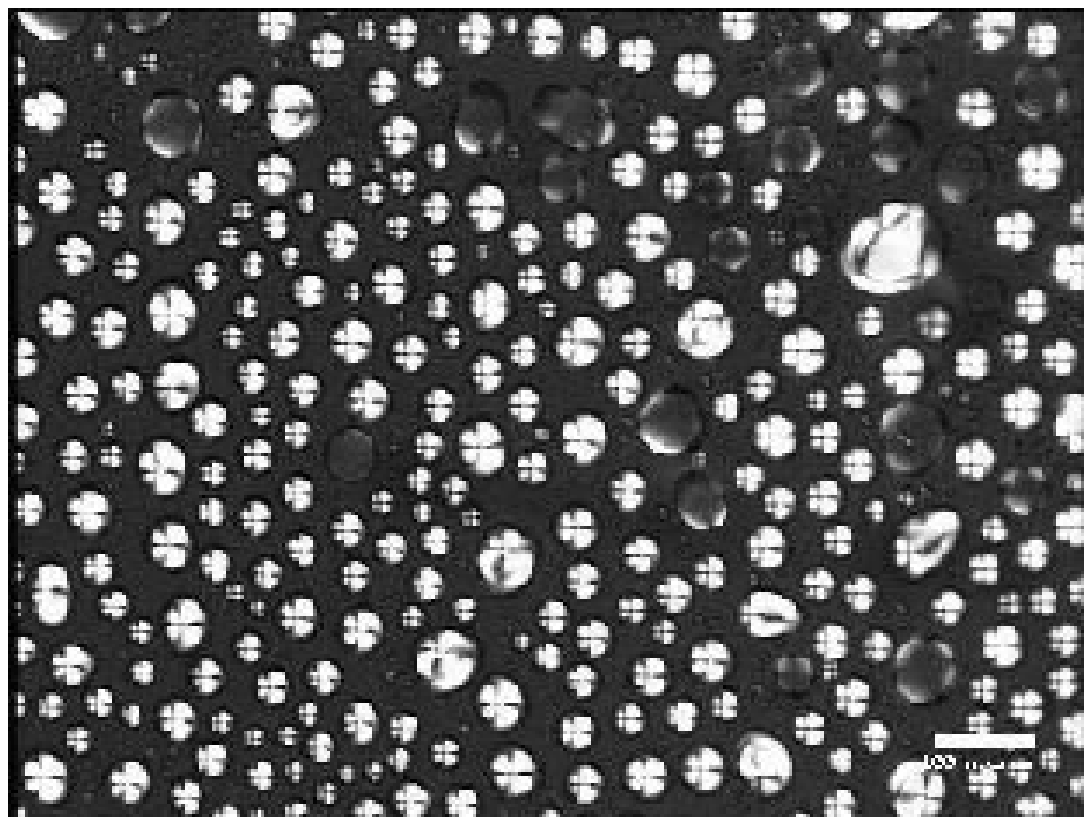,width=0.70\textwidth}}
\subfigure[]{\epsfig{file=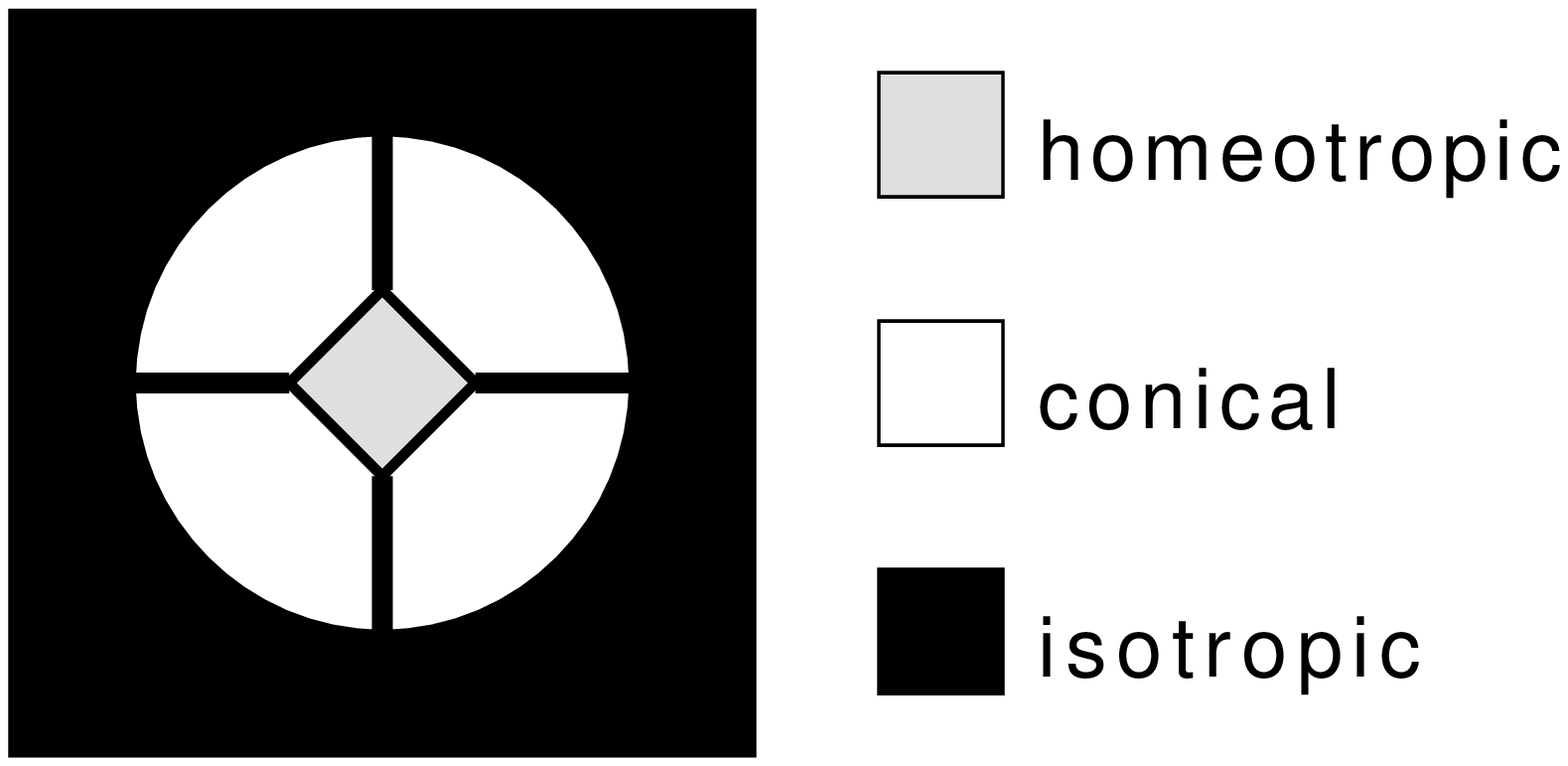,width=0.25\textwidth}}
\caption{
\label{anchtran}
\small{
(a) Nematic to isotropic phase transition in a cell with \textsf{C18}
as aligning layer.
The dark gray background is the homeotropic phase.
In the circular domains the alignment is conical.
The isotropic phase appears inside the domains.
(b) Scheme of the homeotropic to isotropic phase 
transition. 
(c) Isotropic to nematic phase transition in the same cell.
Now the dark background is the isotropic phase and in the 
circular domains the alignment is again conical.
The homeotropic state appears inside the domains.
(d) Scheme of the isotropic to homeotropic phase 
transition. 
}}
\end{center}
\end{figure}

On increasing the temperature, the transition to the isotropic phase
takes place inside the domains (Figures \ref{anchtran}(a) and
\ref{anchtran}(b)).
On cooling from the isotropic phase the bright domains appear
again and the transition to the nematic phase takes place inside the
domains (Figures \ref{anchtran}(c) and
\ref{anchtran}(d)).
%
%\psdraft
%\psfull
%

Following Safran et al.\cite{SafRobGar86}, we think that
the surfactants molecules are arranged in soliton-antisoliton pairs
randomly distributed over the substrate area.
Depending on the length of the tails
and on the temperature, those structures can be
large enough to prevent the uniform homeotropic alignment,
which may be the case in the \textsf{C22} mono-layer.
If the soliton-antisoliton pairs are not too large,
they can give a
uniform homeotropic alignment in the bulk.
However, they are the germs of the conical structures
appearing at the anchoring transition.

\section{Conclusions} 
The surface induced homeotropic alignment in \textsf{NLC} cells 
by \textsf{LB} mono-layers of stearic and behenic acids has been investigated.
Stearic acid was found to be very good for aligning \textsf{NLC}
in the homeotropic state, whereas
behenic does not give a well defined alignment.

A relaxation process from the flow induced quasi-planar orientation to the
surface induced homeotropic orientation has been observed.
It takes place once the cell is filled with the liquid crystal such
that there is not more material flow. 
It starts from the edges of the cell in form of expanding
homeotropic domains.
The disclination lines dividing the homeotropic domains from the quasi-planar
ones have strength $|S| = 1/2$ and move at a speed dependent
on the thickness of the cell.
The speed, which is also the speed with which the homeotropic domains expand
in the cell, was found to decrease as the cell thickness increases,
in accordance with a single model involving both the elastic relaxation of the
liquid crystal splay-bend deformation and the relaxation of the deformed
\textsf{LB} chains.

Further studies of
this relaxation process could be useful to understand the anchoring
mechanism and in estimating anchoring energies.
In the future, computational simulations may play an important r\^ole in
giving answers to several remaining questions, but more experiments are
also needed to formulate the right models.

%The work is continuing with similar investigations using mixed stearic/be- 
%henic mono-layers as aligning films.

\psfull
%

%\clearpage
\bibliography{journals2,LCbibl}
\end{document}